# Unleashing the Potential of Li-Metal Batteries A Breakthrough Ultra-High Room-Temperature Ionic Conductivity Composite Solid-State Electrolyte

Xiong Xiong Liu, Shengfa Feng, Pengcheng Yuan, Yaping Wang, Long Pan*, ZhengMing Sun*

Key Laboratory of Advanced Metallic Materials of Jiangsu Province, School of Materials Science and Engineering, Southeast University, Nanjing 211189, PR China

* Email: zmsun@seu.edu.cn (Z.M.S.); panlong@seu.edu.cn (L.P.)

**Abstract:**

The solid-state electrolyte is critical for achieving next-generation high energy density and high-safety batteries. Solid polymer electrolytes (SPEs) possess great potential for commercial application owing to their compatibility with the existing manufacturing systems. However, unsatisfactory room-temperature ionic conductivity severely limits its application. Herein, an ultra-high room-temperature ionic conductivity composite solid-state electrolyte (CSE) is prepared by introducing an appropriate amount of $SiO_2$ nanosphere to the PVDF-HFP matrix. By doing this, the polymer particles are divided and surrounded by $SiO_2$. And the interface amount is maximized resulting in the high ionic conductivity of 1.35 mS cm$^{-1}$ under room temperature. In addition, the CSE shows a wide electrochemical window of 4.95 V and a moderate Li$^+$ transference number of 0.44. The CSE demonstrates good stability with Li anode, with Li symmetric cells that could cycle 1000 h at a current density of 0.2 mA cm$^{-2}$. The full cell assembled with LiFePO$_4$ (LFP) and Li metal displays a high reversible specific capacity of 157.8 mAh g$^{-1}$ at 0.1C, and it could maintain 92.9% of initial capacity after 300 cycles at 3C. Moreover, the strategy is applied in solid-state sodium/potassium batteries and displays excellent performance.

**Keywords:**

**Introduction**

Solid-state electrolytes are fundamental to solid-state lithium metal batteries that can simultaneously address energy density and safety problems [1-5]. In general, solid-state electrolytes require decent contact with electrodes, wide electrochemical window, and, in particular, high room-temperature ionic conductivity ($10^{-3}$ S cm$^{-1}$) [6,7]. In this sense, solid-state polymer electrolytes (SPEs) have provoked significant research and industrial interest owing to their good flexibility, tight interface with electrodes, and ease of processing, which are compatible with the existing manufacturing systems of lithium-ion batteries [8,9]. However, their applications are impeded by the low ionic conductivity ($10^{-7}$-$10^{-5}$ S cm$^{-1}$ at room temperature) [10].

Incorporating inorganic fillers into SPEs, which forms composite solid-state electrolytes (CSEs), is as an effective approach to enhance the ionic conductivity [11,12]. Generally, active inorganic fillers (AIFs) that are ionic conductive have been added into SPEs, including garnet-type oxide (*e.g.*, Li$_7$La$_3$Zr$_2$O$_{12}$), perovskite-type oxides (*e.g.*, Li$_x$La$_y$TiO$_3$), NASICON-type phosphates (*e.g.*, Li$_{1.3}$Al$_{0.3}$Ti$_{1.7}$(PO$_4$)$_3$), *etc.*, which can reduce the polymer crystallinity and provide additional Li$^+$ pathways [13]. However, AIFs-based CSEs can only deliver ionic conductivities of $10^{-5}$-$10^{-4}$ S cm$^{-1}$ (at room temperature), which is still far from the practical requirement ($10^{-3}$ S cm$^{-1}$). Additionally, other critical challenges such as limited metal resources (Li, Ge, Zr, *etc.*), harsh synthesis conditions, and high production costs, also obstruct the implementation of AIFs [14-16].

Passive inorganic fillers (PIFs) involving SiO$_2$, Al$_2$O$_3$ (*α* phase), TiO$_2$, *etc.*, are more appealing than AIFs when used to build advanced CSEs, as they are cost effective, easy handling, and environmentally compatible, which are essential to large-scale applications. More importantly, PIFs-based CSEs exhibit ionic conductivities of $10^{-5}$-$10^{-4}$ S cm$^{-1}$ (at room temperature), which are comparable to those of AIFs-based CSEs [17-20]. Despite that PIFs are non-lithium-ion (Li$^+$) conductors, they have strong interfacial interactions with Li-salts and can weaken the polymer crystallization, thereby facilitating the Li-salt dissociation and providing more moveable Li$^+$.[12,21-23] Moreover, the interfacial ionic conductivity between inorganic fillers and polymers can reach as high as 10$^{-2}$ S cm$^{-1}$ (at 30 °C) [9,24]. Therefore, maximizing the PIF contents and the PIFs/polymer interfaces promises high-performance PIFs-based CSEs with more moveable Li$^+$ and high room-temperature ionic conductivity.

In this study, a "Polymer in SiO$_2$" (PiSi) composite solid-state electrolyte with ultra-high room-temperature ionic conductivity is developed by introducing SiO$_2$ nanospheres with -OH functional group as passive fillers to poly(vinylidene fluoride-hexafluoropropylene) (PVDF-HFP). With increasing SiO$_2$ content, the morphology of the composite solid-state electrolyte underwent a transition from "SiO$_2$ in Polymer" to "Polymer in SiO$_2$" (Scheme 1). In the PiSi, the SiO$_2$ nanospheres act as separators between the polymer particles, maximizing the interface between PVDF-HFP and SiO$_2$, significantly accelerating the migration of lithium ions and thus improving the ionic conductivity. As a result, the composite electrolyte displayed an ultra-high room temperature ionic conductivity of 1.35 mS cm$^{-1}$ and a wide electrochemical window of 4.95 V (Figure 1a). Additionally, the large size and flexibility of the composite membrane potentially presented practical applications (Figure 1b). The electrolyte exhibited good stability with Li metal, and the Li//Li symmetric batteries showed remarkable long-term cycling stability, working for 1000 h at a current density of 0.2 mA cm$^{-2}$. When assembled with Li metal anode and LiFePO$_4$, the solid-state batteries demonstrated excellent rate performance and cycle stability, retaining 92.9% capacity after 300 cycles at 3C under ambient conditions. Density functional theory (DFT) and molecular dynamics (MD) simulations were also conducted to elucidate the mechanism underlying the significant improvement in ionic conductivity.

The study sheds light on the potential of passive filler-based composite solid-state electrolytes as a promising approach to enhance the performance of SSLMBs.

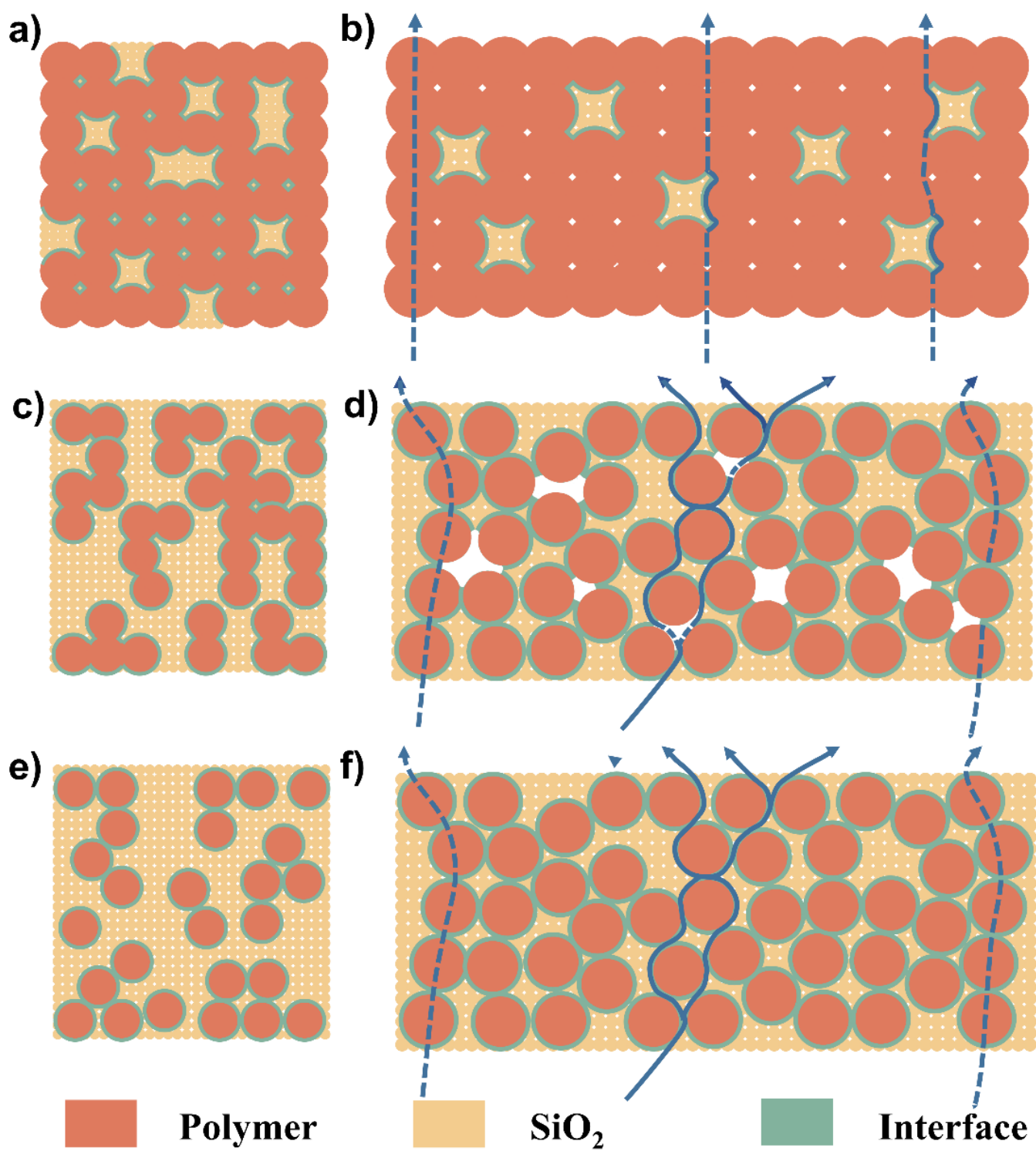

**Scheme 1** (a), (c), and (e) Illustration of composite polymer electrolytes structure from the top of view. (b), (d), and (e) Illustration of Li-ion conduction pathways in composite polymer electrolytes.

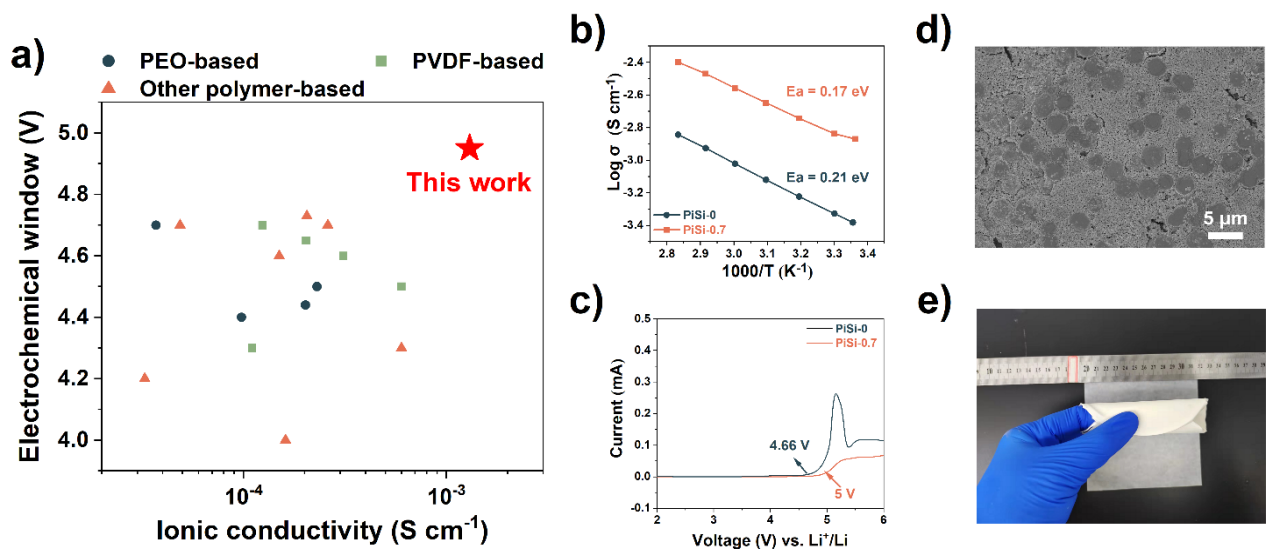

**Figure 1** (a) Electrochemical window is plotted against Li$^+$ ionic conductivity for PiSi. (b) Arrhenius plot for the ionic conductivity of PiSi-0 and PiSi-0.7. (c) LSV curves of PiSi-0 and PiSi-0.7 at a sweeping rate of 0.1 mV s$^{-1}$. (d) SEM image of PiSi-0.7 electrolyte, (e) The digital photo of a solid-state electrolyte film with a diameter of 110 mm.

**Results and Discussion**

The XRD patterns of PiSi-0 and PiSi-0.7 are shown in **Figure S1a**. SiO$_2$ nanoparticles show the amorphous phase, and there are no peaks of SiO$_2$ in the spectrum of PiSi. Both PiSi-0 and PiSi-0.7 show two broad characteristic peaks around 20° and 40° that could be ascribed to the crystalline structure of PVDF-HFP. Furthermore, the crystallinity of PiSi-0.7 (18.69%) is lower than that of PiSi-0 (27.74%), implying more amorphous areas in PiSi-0.7 that could transfer Li-ions faster for higher ionic conductivity. Fourier transform infrared (FT-IR) was conducted to reveal the chemical structure of the SiO$_2$ and polymer electrolytes. As shown in **Figure S1b**, the SiO$_2$ shows four obvious peaks. Among them, the peak at 958 cm$^{-1}$ could be attributed to the characteristic peak of Si-OH [25,26], which can be clearly seen in the spectrum of PiSi-0.7. **Figure S1c** shows the Raman spectra in the frequency range of 720-770 cm$^{-1}$. The peaks at about 741 cm$^{-1}$, 744 cm$^{-1}$, and 749 cm$^{-1}$ are attributed to the free TFSI$^-$ and, CIP (contact ion pairs) and ACG (aggregated ion pairs), respectively [27-29]. After introducing SiO$_2$, the ratio of free TFSI$^-$ peak area increases from 67.1% to 87.4%, which means the increased dissociation of LiTFSI, indicating more free lithium ions in the PiSi-0.7. Therefore, the ionic conductivity could be improved.

The electrochemical stability of PiSi was measured via linear sweeping voltammetry (LSV) under 25 °C (**Figure S2a**). The electrochemical window of PiSi-0.7 is extended from 4.66 V to 5.00 V after the introduction of SiO$_2$ nanoparticles. And the SPE curve shows a slightly broad peak at around 4.2 V, while the current of PiSi-0.7 is nearly zero until 5 V and lower than that of PiSi-0 in the whole voltage range, demonstrating good electrochemical stability. Such an electrochemical window shows excellent application potential in matching high-voltage cathode materials. The lithium transference number ($t_{Li^+}$) was investigated by AC impedance and DC polarization. As shown in **Figure S2b**, the $t_{Li^+}$ value of PiSi-0.7 is 0.44, which is higher than those of PiSi-0 (0.17, **Figure S2c**) and liquid electrolyte (< 0.4) [30,31]. The PiSi-0.7 with a higher Li$^+$ transference number could reduce the anions accumulation during the process of charge/discharge, thereby improving the performance of batteries, especially at high-rate conditions [32,33].

**Figure S3** displays the scanning electron microscopy (SEM) images of the SiO$_2$, and the SiO$_2$ exhibits a uniform spherical morphology with an average diameter of approximately 200 nm. The morphology of the PiSi-0 film depicts a smooth surface and interconnected microspheres, forming a porous structure with visible voids between the polymer particles, as shown in **Figure S4**. **Figure S5** presents the SEM images of the analyzed samples. As the SiO$_2$ content increases, we observe a gradual decrease in the number of polymer particles within the field of view. And it was observed that the polymer particles tend to overlap more in regions where the $SiO_2$ content is low. Specifically, in **Figure S5c**, the voids between polymer particles are not fully filled with $SiO_2$ nanospheres. In the PiSi-0.5 sample, the polymer particles and aggregated $SiO_2$ nanospheres exhibit an intermediate distribution (**Figure S5 d-e**). At this mass ratio, the polymer particles still overlap, and the underlying particles remain visible, indicating that the polymers are not yet entirely separated. Notably, at a mass ratio of 70% $SiO_2$, the polymer particles are immersed in a "sea" of $SiO_2$ (**Figure S5 g-i**). Each polymer particle is surrounded by $SiO_2$, creating a maximal number of interfaces between the two components. Furthermore, upon increasing the weight of $SiO_2$ to 100% of the polymer, the PiSi-1.0 composite film demonstrated difficulty in peeling off from the petri dish, as depicted in **Figure S6**. At the same time, the PiSi-0.7 sample was prepared using 400 nm diameter $SiO_2$. As shown in Figure **S7 a-c**, the SEM images indicated that the nanospheres caused the initial polymer microsphere to tear into irregular strips, resulting in a structure with a low ionic conductivity of 0.12 mS cm$^{-1}$, as demonstrated in **Figure S7d**.

With such a unique structure, the electrolyte shows an ultra-high room temperature ionic conductivity of 1.35 mS cm$^{-1}$ (**Figure S8a**). In comparison, the room-temperature ionic conductivities of PiSi-0, PiSi-0.2, and PiSi-0.5 are just 0.42, 0.57, and 0.92 mS cm$^{-1}$, respectively. In addition, the activation energy of lithium-ion diffusion calculated from the Arrhenius equation is decreased from 0.21 eV to 0.17 eV, which indicates that the Li-ion migration barrier is reduced. The Arrhenius plots of PiSi electrolytes reveal that the ionic conductivities gradually improved with an increasing weight ratio of $SiO_2$. This improvement can be attributed to the growing amount of interface between polymer and $SiO_2$. The corresponding EIS spectra are shown in **Figure S8 b-e**.

The thermal stability of PiSi was evaluated under $N_2$ atmosphere from room temperature to 800 °C,

and the results are shown in **Figure S9**. Despite PiSi-0.7 displaying a lower decomposition temperature (326 °C) than that of PiSi-0 (358 °C), the PiSi-0.7 is stable until 300 °C. The slight weight loss observed below 100 °C could be attributed to the moisture evaporation. The introduction of $SiO_2$ significantly reduced the trapped moisture content from 8.5% to about 0%, implying improved water stability of the PiSi-0. The contents of the residual solvent of PiSi-0.7 and PiSi-0 are 2.9% and 3.3%, respectively. The similar content of DMF has almost the same effect on the two samples, indicating that the reason for the ultra-high ionic conductivity of PiSi-0.7 is not the residual solvent.

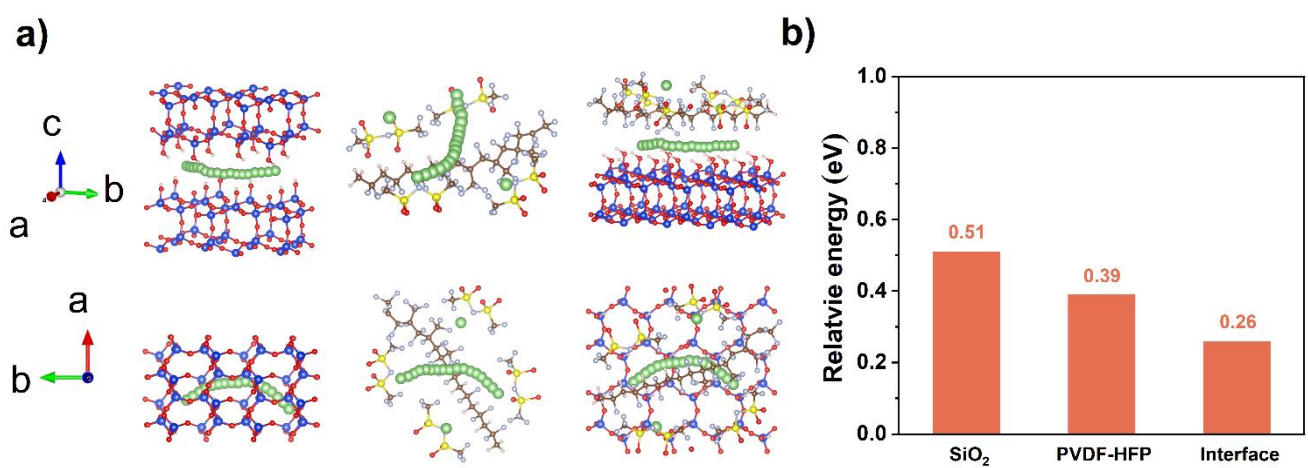

**Figure 2** (a) The pathway of $Li^+$ migration in interface between $SiO_2$ and $SiO_2$, PVDF-HFP, and interface between $SiO_2$ and PVDF-HFP. (b) DFT calculation results of Li diffusion along interface between $SiO_2$ and $SiO_2$, PVDF-HFP and interface between $SiO_2$ and PVDF-HFP.

The mechanism underlying the enhancement of ionic conductivity was further elucidated through density functional theory (DFT), solid-state nuclear magnetic resonance (SSNMR), and molecular dynamics (MD) simulations. We first calculated the Li-ion migration barriers along pure polymer chains, $SiO_2$, the interface between $SiO_2$ and $SiO_2$, and the interface between polymer and $SiO_2$ by DFT (**Figure 2**). The results show that the migration barrier along the interface between $SiO_2$ and $SiO_2$ is 0.51 eV which is significantly higher than that along the polymer chain (0.39 eV), suggesting the Li$^+$ might not be transported in the aggregated SiO$_2$. Conversely, the barrier along the interface between the polymer and SiO$_2$ exhibits the lowest value (0.26 eV), implying that it is the most accessible pathway for the migration of Li$^+$. We also calculated the migration barrier of Li$^+$ from the polymer phase transfer to the surface of SiO$_2$ and back to the polymer phase (**Figure S10**). The results revealed that the migration barriers of transfer between the polymer phase and SiO$_2$ phase (0.30 eV) are even lower than the value along the polymer chain. Therefore, it is speculated that the lithium ions in polymer tend to transfer to the interface between polymer and SiO$_2$ and migrate in the continuous interface "highway" provided by the structure of "Polymer in SiO$_2$".

To demonstrate the significance of the interface between polymer and SiO$_2$, we performed solid-state NMR (SSNMR) analysis to investigate the Li$^+$ migration behavior in Li symmetric cells with $^6$Li metal electrodes (**Figure 3a** and **Figure S11a**). We first conducted the SSNMR on the LiTFSI and LiTFSI mixed with SiO$_2$, and the results are presented in **Figure S11b**. The $^6$Li resonances of lithium salt and lithium salt mixed with SiO$_2$ are observed at about -0.656 and 0.667 ppm, respectively. The SSNMR spectra of PiSi-0 and PiSi-0.7 are displayed in **Figure S11c**. The $^6$Li resonance at 0.03 ppm belongs to the Li$^+$ in the PVDF-HFP matrix, meaning the Li-ions are not transported by SiO$_2$, consistent with the DFT calculation results. With the addition of SiO$_2$, a new peak at around -0.03 ppm appears, corresponding to Li$^+$ at the interface between the polymer and SiO$_2$. The SSNMR spectra of pristine PiSi-0.7 and cycled PiSi-0.7 are shown in **Figure 3b**. The peak area of Li$^+$ at the interface is obviously increased, as demonstrated by the quantitative analysis in **Figure 3c**. Li$^+$ form PVDF-HFP and interface account for 99.0% and 1.0%, respectively, before $^6$Li cycling. The $^6$Li amount at interfaces is increased to 53.1%, nearly 53-fold, while the $^6$Li amount in polymers is decreased from

99.0% to 46.9%. The SSNMR result suggests that the Li$^+$ transport mainly depends on the interfaces in the "polymer in SiO$_2$" structure.

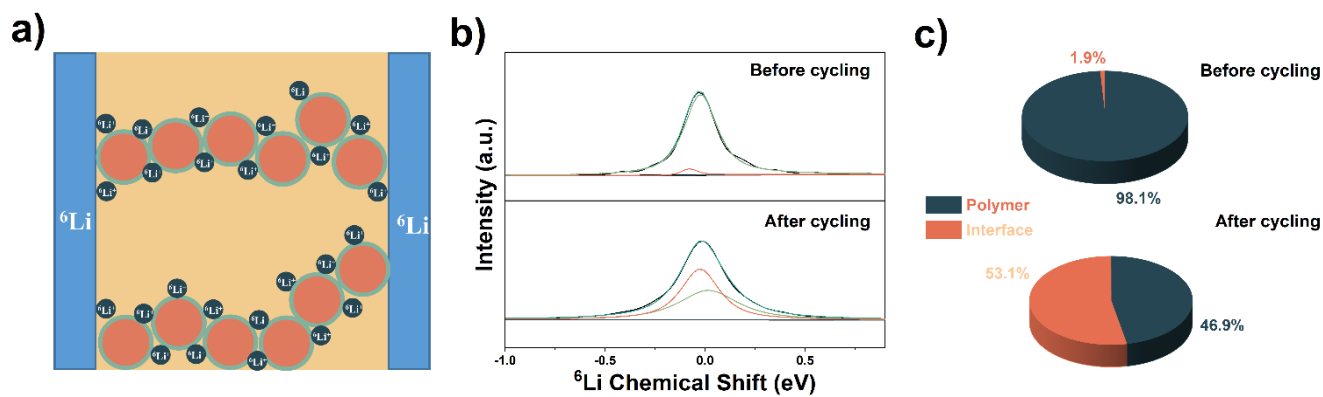

**Figure 3** (a) Li$^+$ migration behavior with PiSi-0.7. (b) $^6$Li SSNMR spectra of PiSi-0.7 before and after cycling (c) Quantitative analysis of $^6$Li amount in PVDF-HFP and the interface between PVDF-HFP and SiO$_2$.

Furthermore, MD simulations were employed to investigate the Li$^+$ transport behavior on a larger scale. In our simulations, polymer, "SiO$_2$ in polymer (PiSi-0.2)", and "polymer in SiO$_2$ (PiSi-0.7)" are compared to identify differences in Li$^+$ migration. **Figure 4a** shows the snapshots of the three systems. The mean squared displacements (MSD) were calculated and displayed in **Figure 4b**. After introducing SiO$_2$, the MSD curve of composite solid electrolytes is steeper than the pure polymer electrolyte, implying addition of SiO$_2$ could accelerate ion migration. Furthermore, the MSD curve of PiSi-0.7 system shows the steepest slope, suggesting the fastest ion transport. The corresponding diffusion coefficients ($D_{Li^+}$) for each system are estimated to be $1.22 \times 10^{-9}$ cm$^2$ s$^{-1}$, $1.30 \times 10^{-9}$ cm$^2$ s$^{-1}$, and $2.17 \times 10^{-9}$ cm$^2$ s$^{-1}$ based on the MSD between 10 ns and 18 ns. The $D_{Li^+}$ was increased to 106% and 178% by introducing SiO$_2$. The coordination number (CN) results are shown in

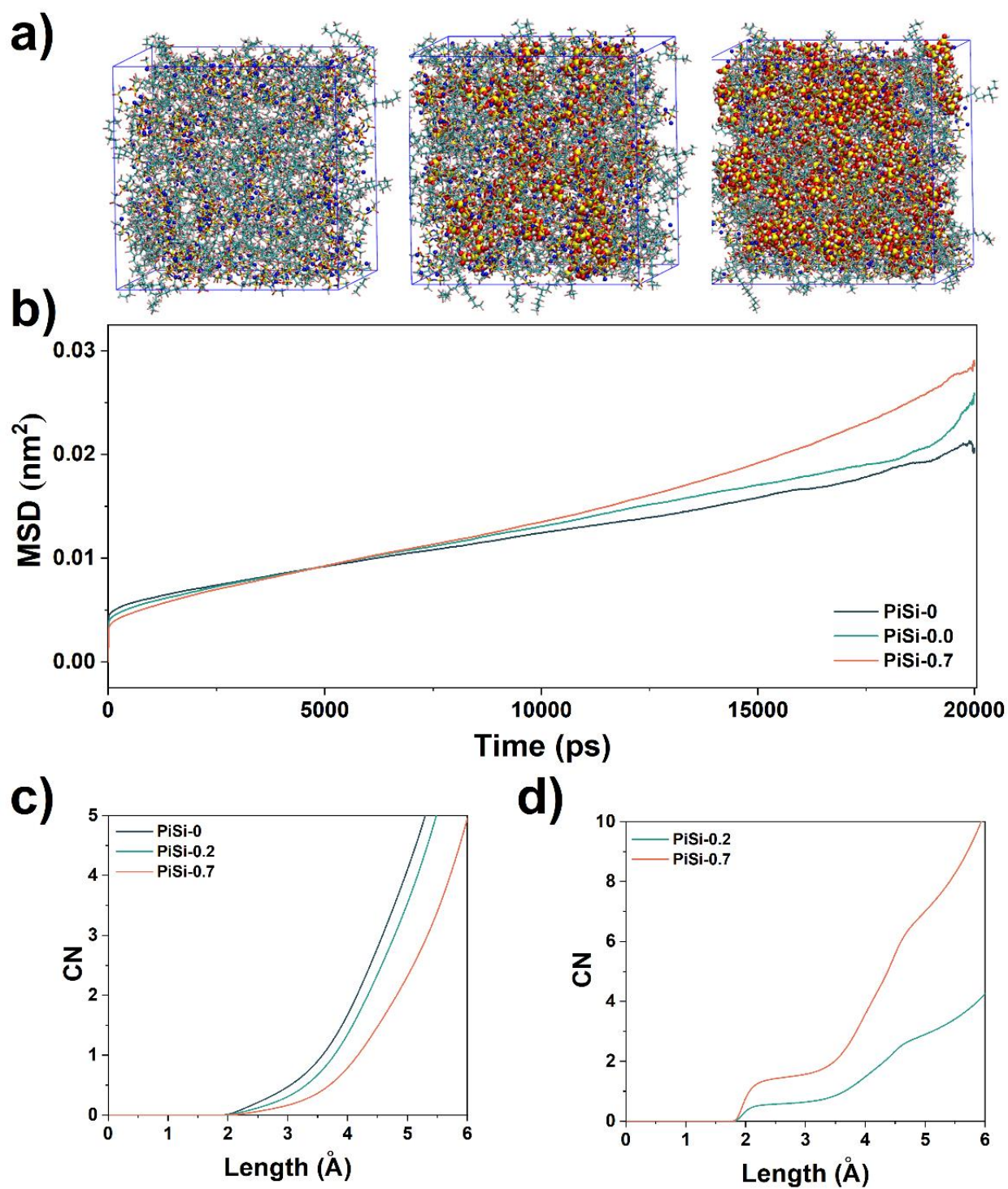

**Figure 4** (a) Snapshot from MD simulation of PiSi-0, PiSi-0.2, and PiSi-0.7. (b) MSD of Li$^+$ in PiSi-0, PiSi-0.2, and PiSi-0.7. (c) Coordination number (CN) of F and (d) Coordination number (CN) of O with reference to Li$^+$ in PiSi-0, PiSi-0.2, and PiSi-0.7.

**Figure 4c and d**, and the corresponding radial distribution functions (g(r)) are shown in **Figure S12a and b**, respectively. **Figure 4c** demonstrates the PVDF-HFP had a weaker correlation with Li-ion as the amount of $SiO_2$ increased, while the O atom of $SiO_2$ had a more robust interaction with Li-ion (**Figure 4d**), indicating that the $Li^+$ tended to transfer to the interface between PVDF-HFP and $SiO_2$. The remarkable enhancement of ionic conductivity by "polymer in $SiO_2$" could be attributed to a large amount of interface in the system, consistent with the conclusion of DFT simulations and SSNMR results.

The interfacial stability of SPE with Li metal was investigated by Li‖PiSi-0.7‖Li symmetric cells at 25 °C. The critical current density (CCD) was determined first under a fixed Li plating/stripping time of 30 min with a step increase in current density of 0.05 mA $cm^{-2}$, and the results are shown in **Figure S13a**. The PiSi-0.7 film exhibited a higher CCD of 0.70 mA $cm^{-2}$ than the PiSi-0 (0.40 mA $cm^{-2}$). The long-term stability of PiSi-0.7 electrolyte was further evaluated at various current densities ranging from 0.05 to 0.2 mA $cm^{-2}$ with 2 h each cycle. The symmetric cells displayed a low overpotential of 53.5, 84.6, 96.1, and 104.1 mV at 0.05, 0.10, 0.15, and 0.20 mA $cm^{-2}$, respectively. And the overpotential showed a negligible increase after 1000 h of Li plating/stripping, as shown in **Figure S13b**. Without $SiO_2$, the symmetric cell failed rapidly after 50 h and 30 h at a current density of 0.1 mA $cm^{-2}$ and 0.2 mA $cm^{-2}$, respectively, indicating poor cycling stability (**Figure S13c and d**).

With such a high ionic conductivity and good compatibility with Li metal, the PiSi-0.7 film was assembled into a full cell with $LiFePO_4$ and Li (LFP‖PiSi-0.7‖Li) to evaluate the practical application potential. As shown in **Figure 5a**, the full cell exhibited remarkable rate performance at 25 °C, delivering high capacities of 157.8, 154.3, 148.1, 139.6, and 126.1 mAh $g^{-1}$ at 0.1, 0.2, 0.5, 1, and 2C, respectively. Even at a high rate of 5C, the full cell can exhibit 75.6 mAh $g^{-1}$. In addition, the capacity

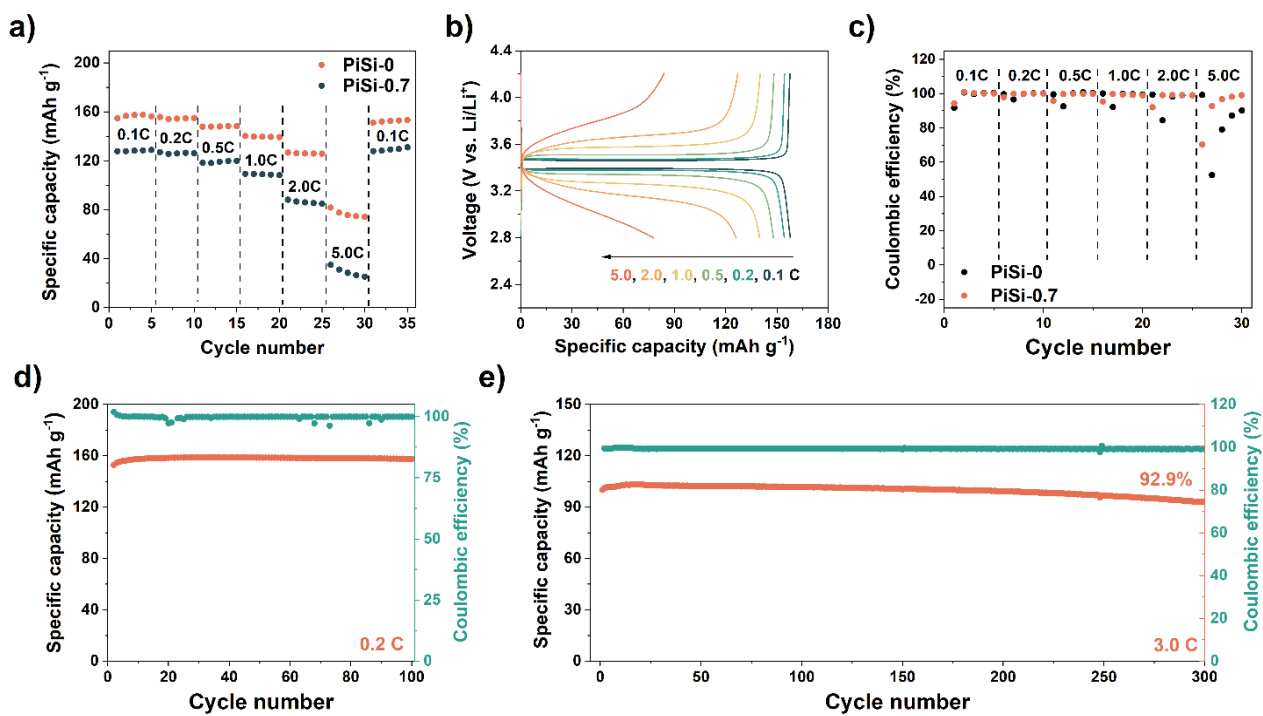

**Figure 5** (a) Rate capability of LFP||PiSi-0||Li and LFP||PiSi-0.7||Li cells at various current rates at 25 °C. (b) Discharging-charging profiles of LFP||PiSi-0.7||Li cells (c) Coulombic efficiencies of LFP||PiSi-0||Li and LFP||PiSi-0.7||Li cells at various current rates at 25 °C. Long-term cycling performance of LFP||PiSi||Li at (d) 0.2 C and (e) at 3C at 25 °C.

maintained its initial value when the rate was changed to 0.1C, demonstrating its good rate performance. In contrast, without the SiO$_2$, the LFP||PiSi-0||Li shows lower capacities at corresponding rates. Even at a current rate of 2C, the discharge specific capacity of LFP||PiSi-0||Li is only 80 mAh g$^{-1}$, close to the capacity of LFP||PiSi-0.7||Li at 5C. The capacity rapidly decreased to 20 mAh g$^{-1}$ when the current rate increased to 5C. **Figure 5b and Figure S13a** show the charge-discharge profiles of LFP||PiSi-0.7||Li and LFP||PiSi-0||Li cells at various current rates, respectively. The LFP||PiSi-0.7||Li cell exhibited much smaller polarization voltages than the LFP||PiSi-0||Li cell. In addition, the LFP||PiSi-0.7||Li cell exhibited a high and stable Coulombic efficiency of ~99% at various current

rates, as **Figure 5c** shows. The cycling performances of LFP||PiSi-0.7||Li at 0.2C and 3C under 25 °C are shown in **Figure 5d and e**, respectively. After 100 and 300 cycles, the LFP||PiSi-0||Li retains 100% and 92.9% of its initial capacity with an average coulombic efficiency (CE) of 99.8% and 99.5%, respectively. Furthermore, when cycled at 5C, the LFP||PiSi-0.7||Li cell shows 77.5% capacity retention after 300 cycles (**Figure S14 b**). Moreover, the full cell can deliver a high capacity of 137.9 mAh g$^{-1}$ (corresponding to 81.9% of initial capacity) after 800 cycles at 2C under 60 °C (**Figure S15 b**).

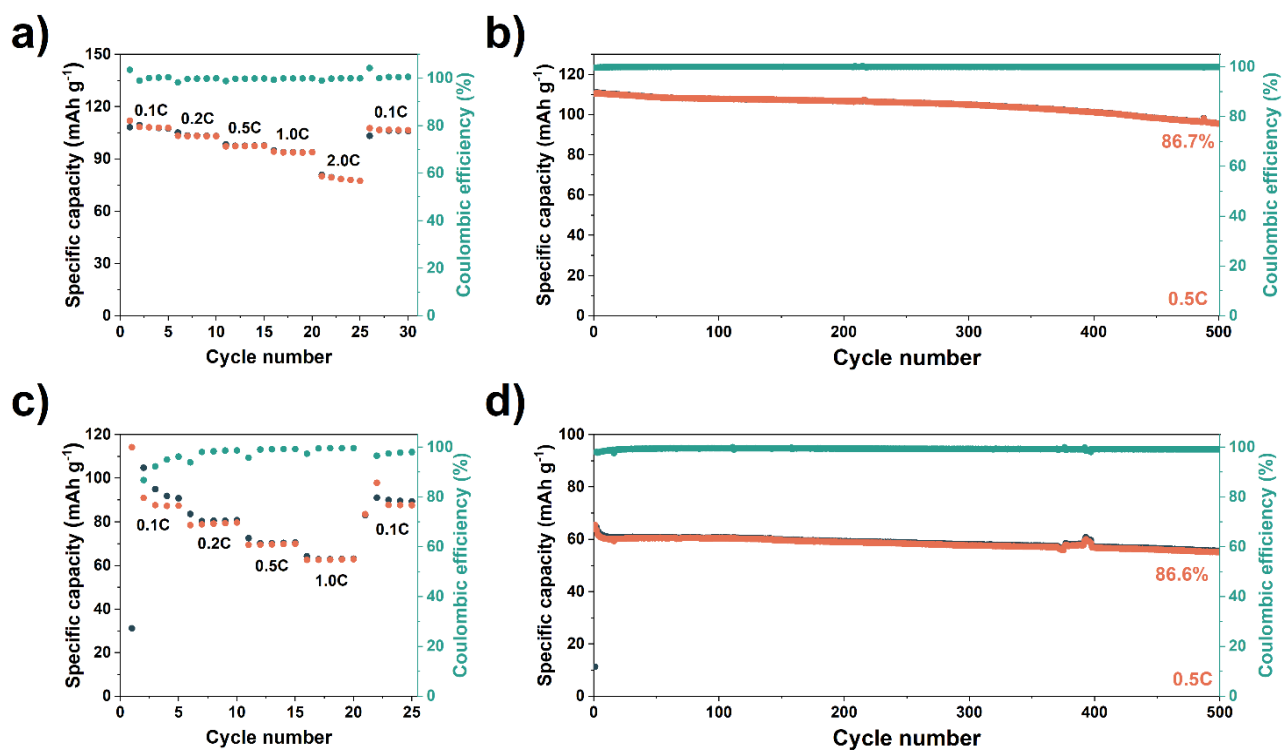

**Figure 6** Rate capability of (a) NVP||PiSi-0.7||Na and (b) KPB||PiSi-0.7||K at various current rates at 25 °C. Long-term cycling performance of NVP||PiSi-0.7||Na and (b) KPB||PiSi-0.7||K at 0.5C at 25 °C

Inspired by the excellent solid-state lithium metal batteries performance, we tested the performance of electrolyte film in solid-state sodium and solid-state potassium metal batteries. **Figure 6a** shows the rate performance of NVP||PiSi-0.7||Na. The cell can deliver a high capacity of 108.0 and

78.5 mAh g$^{-1}$ at current densities of 0.1C and 2C, respectively. When cycled at 0.5C, a high reversible capacity of 95.5 mAh g$^{-1}$ after 500 cycles (86.7% of initial capacity) remained, demonstrating excellent stability (**Figure 6b**). The KPB||PiSi-0.7||K display specific capacities of 87.7 and 629 mAh g$^{-1}$ at 0.1C and 1.0C, respectively (**Figure 6c**). The long-term cycling performance is shown in **Figure 6d**. After 500 charge-discharge cycles at 0.5C, the capacity could maintain 86.6% of the first cycle value, indicating the outstanding stability of PiSi-0.7.

**Conclusion**

In this study, SiO$_2$ nanospheres with a diameter of 200 nm were introduced to the PVDF-HFP to form a "polymer in SiO$_2$" structure composite solid-state electrolyte. The originally aggregated polymer particles were separated into individual particles by SiO$_2$ spheres. Benefiting from the unique structure, the electrolyte shows ultra-high ionic conductivity of 1.35 mS cm$^{-1}$ and 3.99 mS cm$^{-1}$ at 25 °C and 80 °C, respectively. The SSNMR, DFT, and MD simulations reveal the reason for such a high ionic conductivity: the large amount of interface between polymer and SiO$_2$ offers major pathways for Li$^+$ transport. In addition, the electrolyte demonstrated a wide electrochemical window, a moderate Li$^+$ transference number, and a high CCD. When assembled with LFP and Li metal, the LFP||PiSi-0.7||Li full cell delivers a high specific capacity of 157.8 mAh g$^{-1}$ and 75.6 mAh g$^{-1}$ at 0.1C and 5C, respectively. Furthermore, the full cell showed stable cycling stability (92.9% capacity retention) at 3C after 300 cycles under 25 °C. At last, the PiSi structure is successfully applied to the solid-state sodium and potassium batteries, proving the universal of this strategy for alkali metal batteries. The "polymer in SiO$_2$" structure is a promising strategy for fabricating composite solid-state electrolytes with high room-temperature ionic conductivity for SSLMBs.

**Acknowledgment**

The authors appreciate the financial funding from National Natural Science Foundation of China (No. 52201242 and 52250010), Natural Science Foundation of Jiangsu Province (No. BK20200386), Young Elite Scientists Sponsorship Program by CAST (No. 2021QNRC001), and the Fundamental Research Funds for the Central Universities.

# Unleashing the Potential of Li-Metal Batteries A Breakthrough Ultra-High Room-Temperature Ionic Conductivity Composite Solid-State Electrolyte

Xiong Xiong Liu, Shengfa Feng, Pengcheng Yuan, Yaping Wang, Long Pan*, ZhengMing Sun*

Key Laboratory of Advanced Metallic Materials of Jiangsu Province, School of Materials Science and Engineering, Southeast University, Nanjing 211189, PR China

* Email: zmsun@seu.edu.cn (Z.M.S.); panlong@seu.edu.cn (L.P.)

**Experimental Section**

*Synthesis of SiO₂.* SiO$_2$ was synthesized according to a reported method. Typically, 225 mL of ethanol, 30 mL of distilled water, and 5 mL of NH$_3$·H$_2$O (99.0%, Macklin) were mixed and stirred for 30 min. Then, 15 mL of tetraethyl orthosilicate (TEOS, 99%, Aladdin) was added to the solution. After reaction for 2 h at room temperature, the white precipitate was filtrated and washed with ethanol three times and then dried at 80 °C overnight under vacuum.

*Preparation of PiSi.* The PiSi film was prepared by a simple solution-cast method. First, 0.6 g of PVDF-HFP (M.W. = 455000, Macklin) and 0.6 g of LiTFSI (99.9%, Macklin) were dissolved in 6 mL DMF and kept stirred until a homogenous solution was obtained. Then, different amount of (0, 20 wt%, 50 wt%, 70 wt%, 100 wt% for PVDF-HFP) as-prepared SiO$_2$ nanoparticle was added to the solution and stirred for another 12 h. Subsequently, the solution was transferred into a petri dish. The petri dish was first dried in an oven to remove most of the solvent and then transferred into a vacuum oven at 120 °C for 24 h. After the temperature was cooled down, the film was stored in the glove box. The samples were denoted as PiSi-0, PiSi-0.2, PiSi-0.5, PiSi-0.7 and PiSi-1.0. In addition, the composite film with 70 wt% of 400 nm SiO$_2$ for PVDF-HFP was prepared using the same method. The PiSi film for Na and K solid-state batteries is prepared using the same method except that the LiTFSI is replaced by NaTFSI and KTFSI, respectively.

*Material Characterizations.* The crystallographic information was confirmed by X-ray diffractometer by a Haoyuan DX-2700BH diffractor (Cu Kα, λ = 1.5418 Å). The morphology of the sample was studied by FEI Sirion field emission SEM (FESEM). The XPS profiles were obtained by a Thermo K-Alpha spectrometer. The FT-IR and Raman spectra were collected by xxx and xxx, respectively. The SSNMR was conducted by xxx.

*Electrochemical tests.* CR 2032-type coin cells were used to test the performance of solid-state electrolyte, Li symmetric cells, and full batteries. The ionic conductivity was measured by EIS at a frequency range from $10^{-1}$ to $10^6$ Hz and an applied amplitude of 5 mV in symmetric cells of stainless steel (SS)||SSE||stainless steel (SS). The electrochemical window of PiC was tested in SS||SSE||Li cells by LSV at a scan rate of 0.1 mV s$^{-1}$ from 1 to 6 V. The Li+ transference number was tested by DC polarization and AC impedance in Li||SSE||Li cells. For full batteries, LiFePO$_4$ (LFP) was used as cathode material. The slurry containing 80 wt% of LFP, 10 wt% of super P, and 10 wt% of PVDF was cast on Al foil, and dried under vacuum at 120 °C for 12 h. The mass loading of active materials is 1.5~2 mg cm$^{-2}$. The Na$_3$V$_2$(PO$_4$)$_3$ (NVP) and potassium Prussian blue (KPB) were prepared using the same methods as LFP. To ensure good Li$^+$ transport in the cathode and good interfacial contact between LFP and solid-state electrolyte, 5 μL of liquid electrolyte of (1 M LiPF$_6$ in EC/DEC (1:1, v/v)) was added to the cathode side.

**Computational methods.**

The Vienna Ab Initio Package (VASP) was employed to perform all the density functional theory (DFT) calculations within the generalized gradient approximation (GGA) using the Perdew, Burke, and Enzerhof (PBE) formulation.[1-3] The projected augmented wave (PAW) potentials were applied to describe the ionic cores and take valence electrons into account using a plane wave basis set with a kinetic energy cutoff of 450 eV.[4,5] Partial occupancies of the Kohn–Sham orbitals were allowed using the Gaussian smearing method and a width of 0.05 eV. The electronic energy was considered self-consistent when the energy change was smaller than $10^{-4}$ eV. A geometry optimization was considered convergent when the force change was smaller than 0.05 eV/Å. Grimme's DFT-D3 methodology was used to describe the dispersion interactions.[6] The equilibrium lattice constants of structures were optimized when using a 1×1×1 Monkhorst-Pack k-point grid for Brillouin zone sampling. The Climbing Image-Nudged Elastic Band methods had been employed to calculate the Li ions migration barriers in the $SiO_2$ Slab, $SiO_2$-$SiO_2$, PVDF-HFP + LiTFSI with $SiO_2$ slab structures.

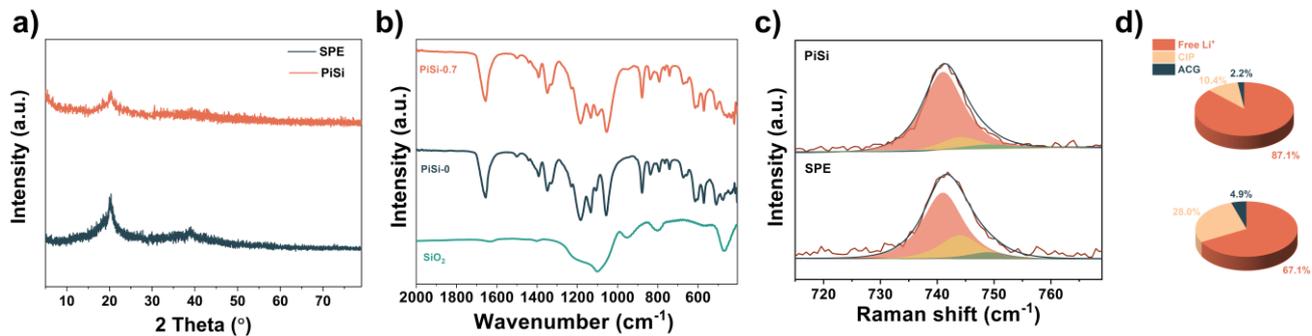

**Figure S1** (a) XRD spectra, (b) FT-IR spectra, (c) Raman spectra of PiSi-0 and PiSi-0.7, (d).

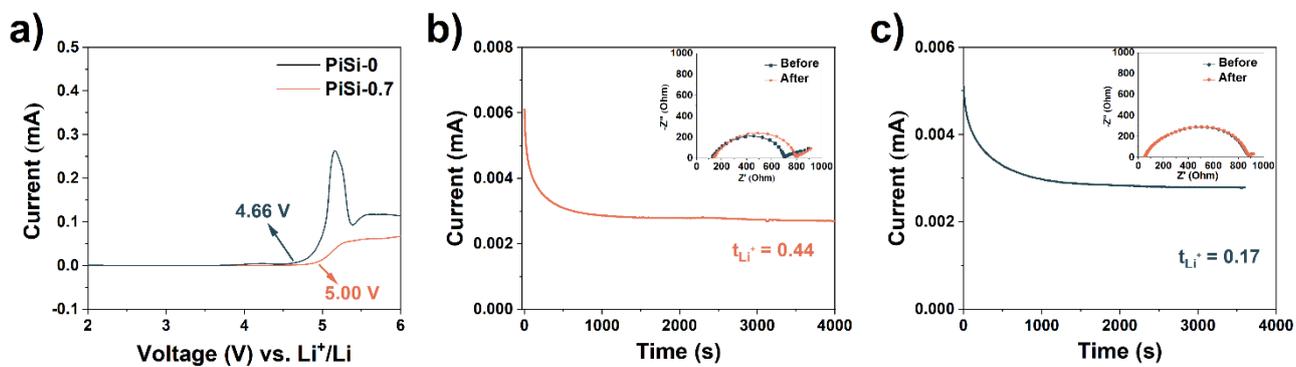

**Figure S2** (a) LSV curves of PiSi-0 and PiSi-0.7 at a sweeping rate of 0.1 mV s$^{-1}$ under 25 °C. Li$^+$ transference number measurement of (b) PiSi-0.7 and (c) PiSi-0 with a DC polarization voltage of 15 mV at 25 °C. Inset: AC impedance spectra before and after polarization respectively.

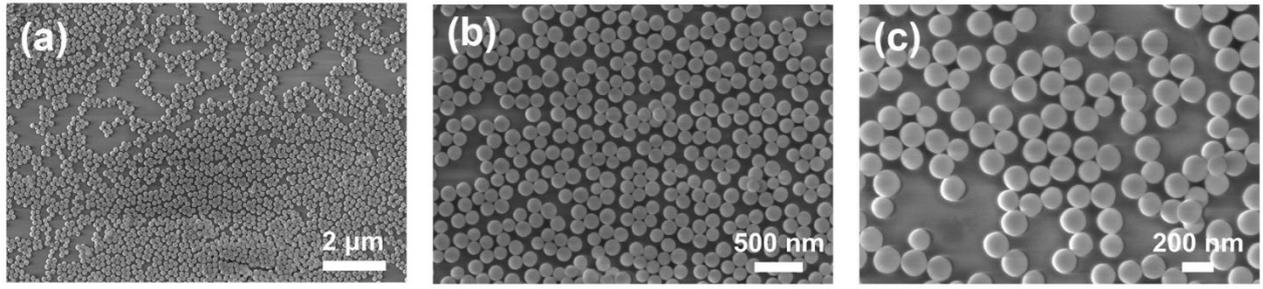

Figure S3 SEM images of SiO$_2$ nanospheres

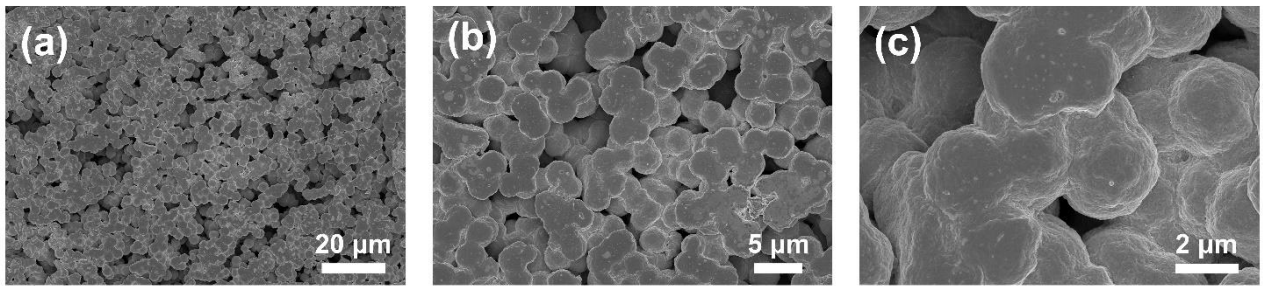

Figure S4 (a)-(c) SEM images of PiSi-0.

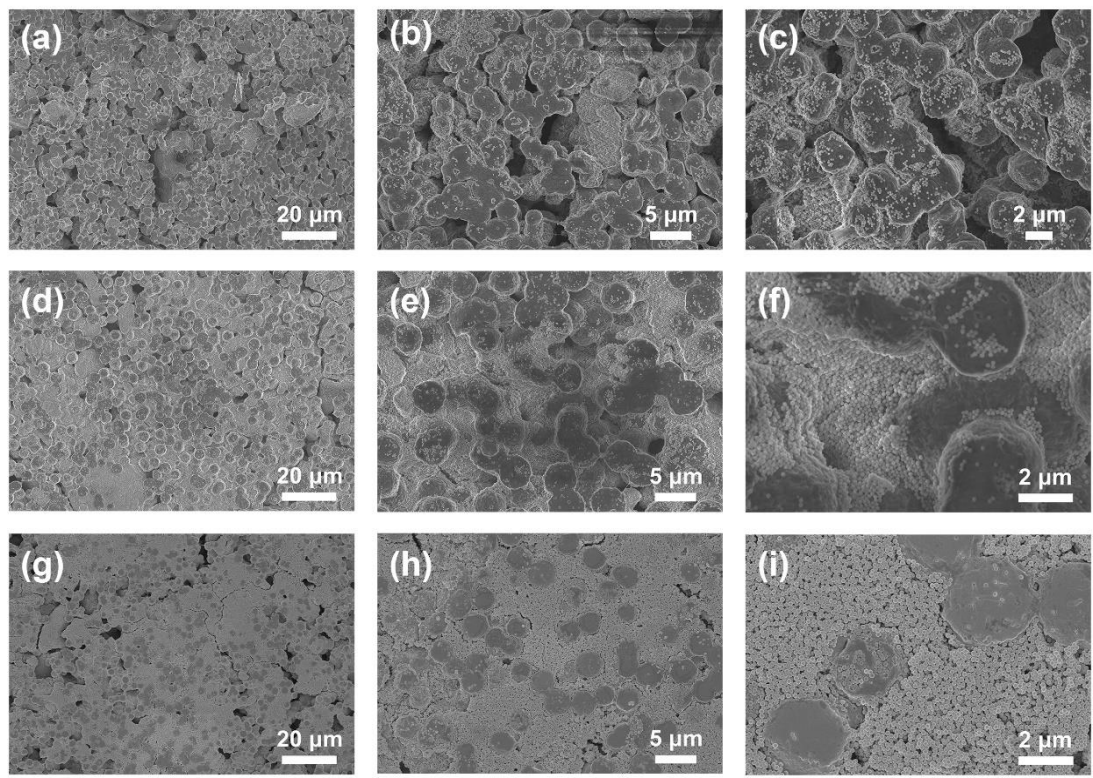

**Figure S5** SEM images of (a)-(c) PiSi-0.2, (d-f) PiSi-0.5, (g-i) PiSi-0.7.

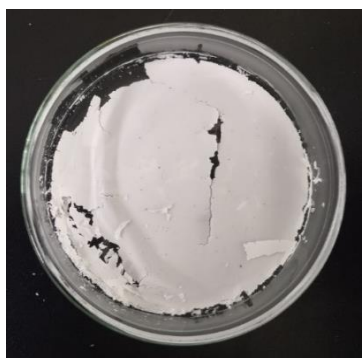

**Figure S6** Digital photo of PiSi-1.0.

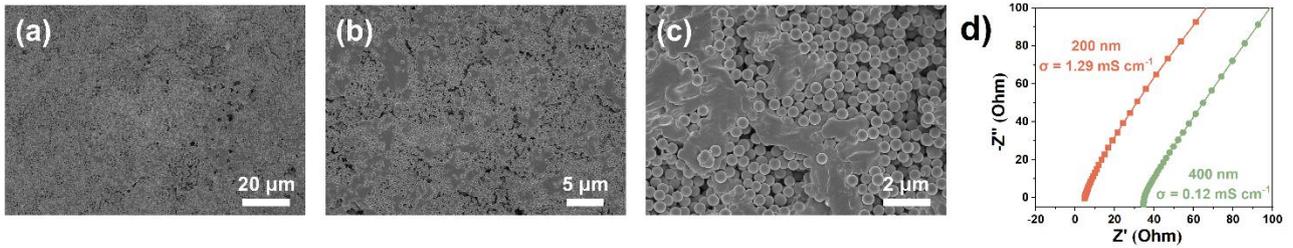

**Figure S7** (a)-(c) Top view SEM images of PiSi-0.7 with 400 nm SiO$_2$. (d) Impedance spectra of PiSi-0.7 with 200 nm and 400 nm SiO$_2$ nanospheres.

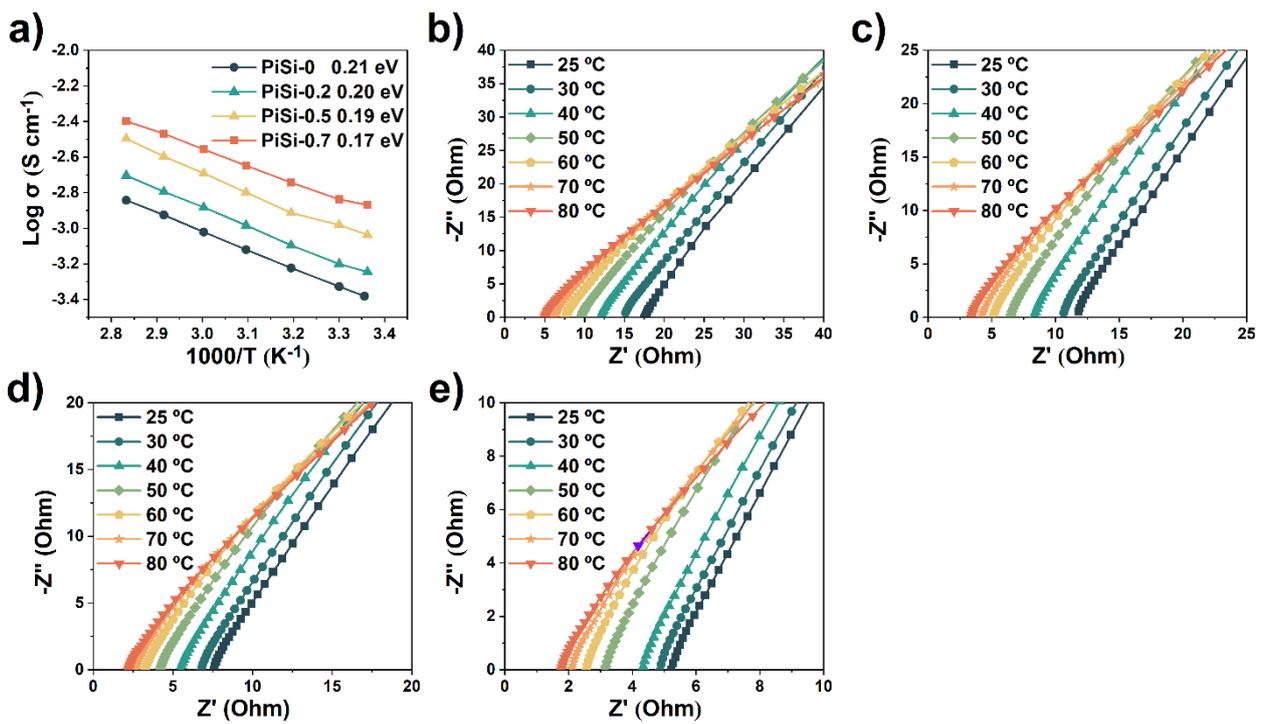

**Figure S8** (a) Arrhenius plot for the ionic conductivity of PiSi samples. (b)-(e) Corresponding impedance spectra of PiSi samples

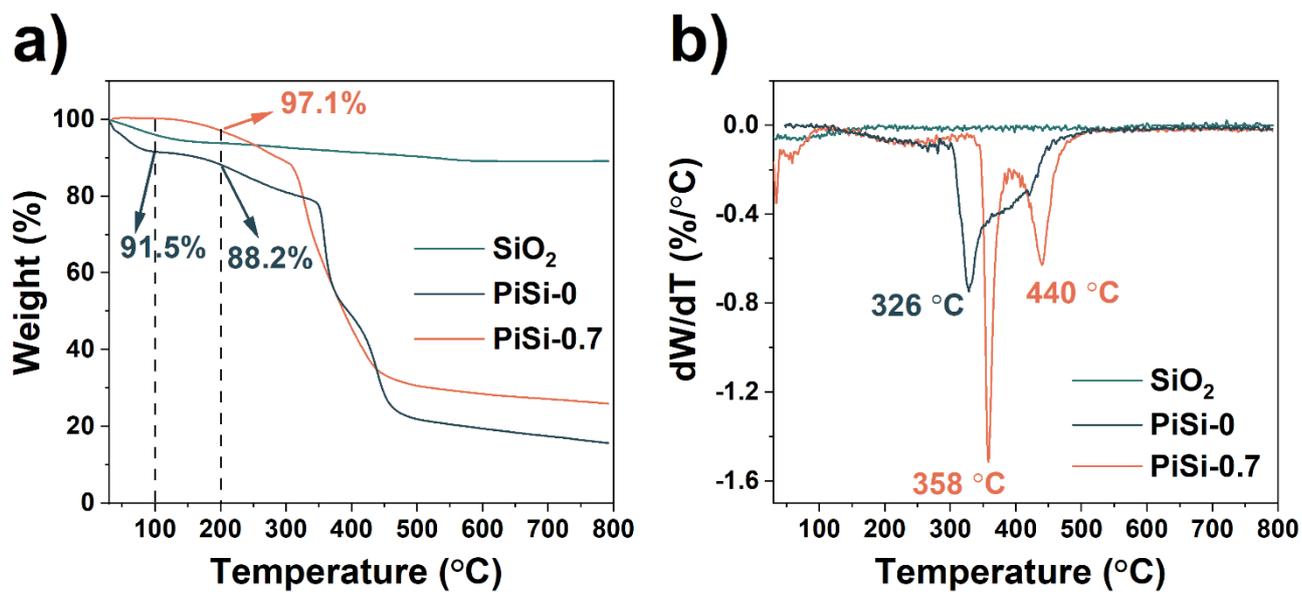

**Figure S9** (a) TG and (b) DTG curves of SiO$_2$, PiSi-0, and PiSi-0.7.

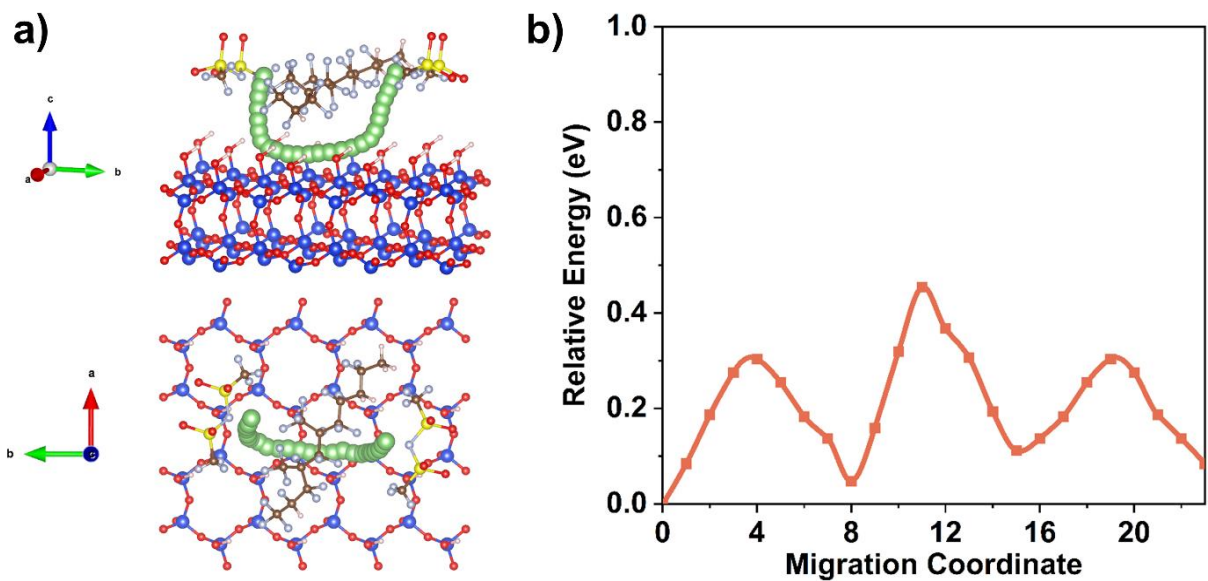

**Figure S10** (a) The pathway and (b) Relative energy of Li+ migration from SiO2 to PVDFH-HFP, PVDF-HFP to SiO2.

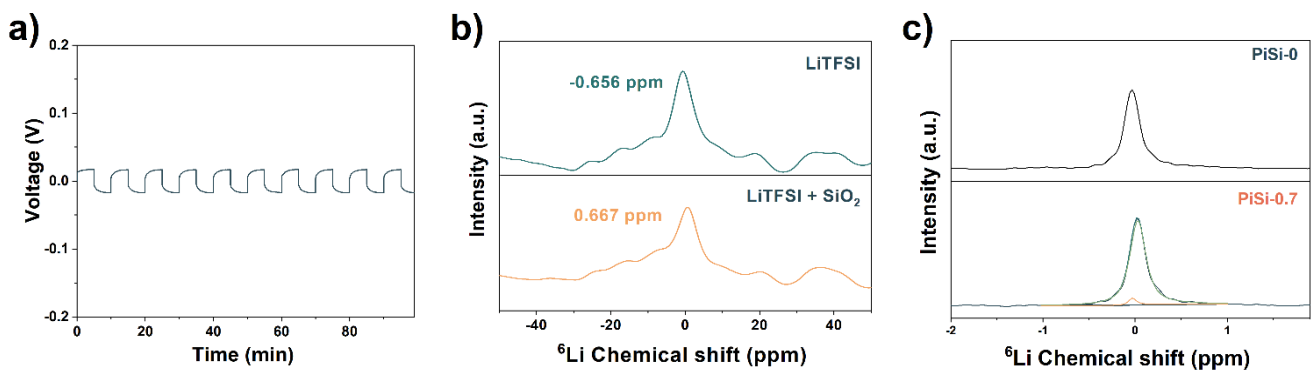

**Figure S11** (a) Li stripping and plating curve at a constant current of 50 μA for 6Li symmetric cells. (b) 6Li SSNMR spectra of LiTFSI and LiTFSI + SiO$_2$ (c) 6Li SSNMR spectra of PiSi-0 and PiSi-0.7 before cycling.

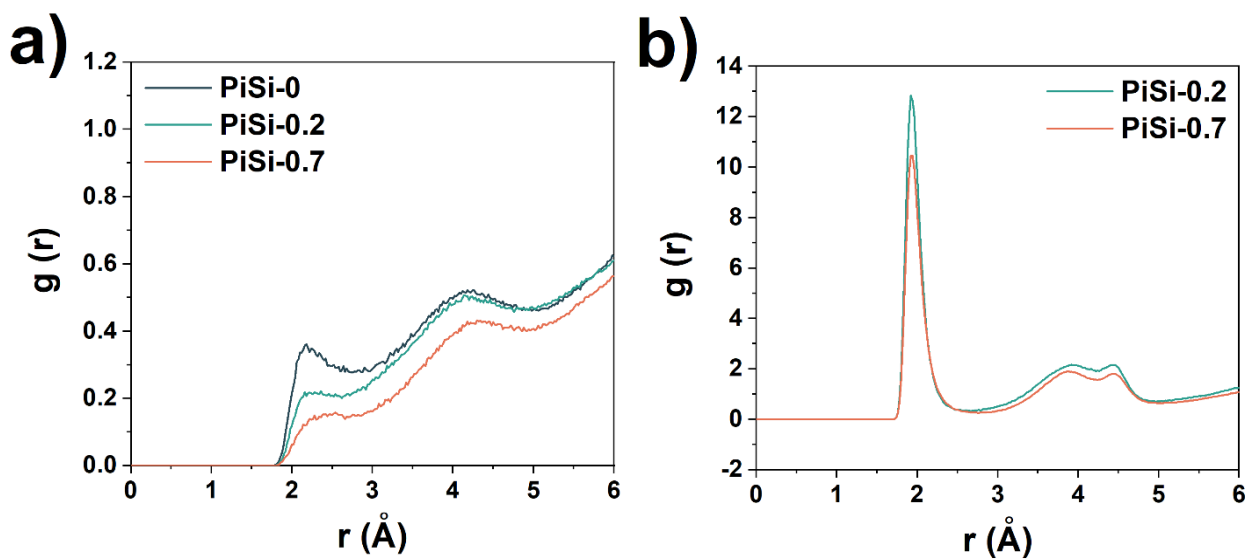

**Figure S12** (a) Radial distribution function (g(r)) of F and (b) Radial distribution function (g(r)) of O with reference to Li$^+$ in PiSi-0, PiSi-0.2, and PiSi-0.7.

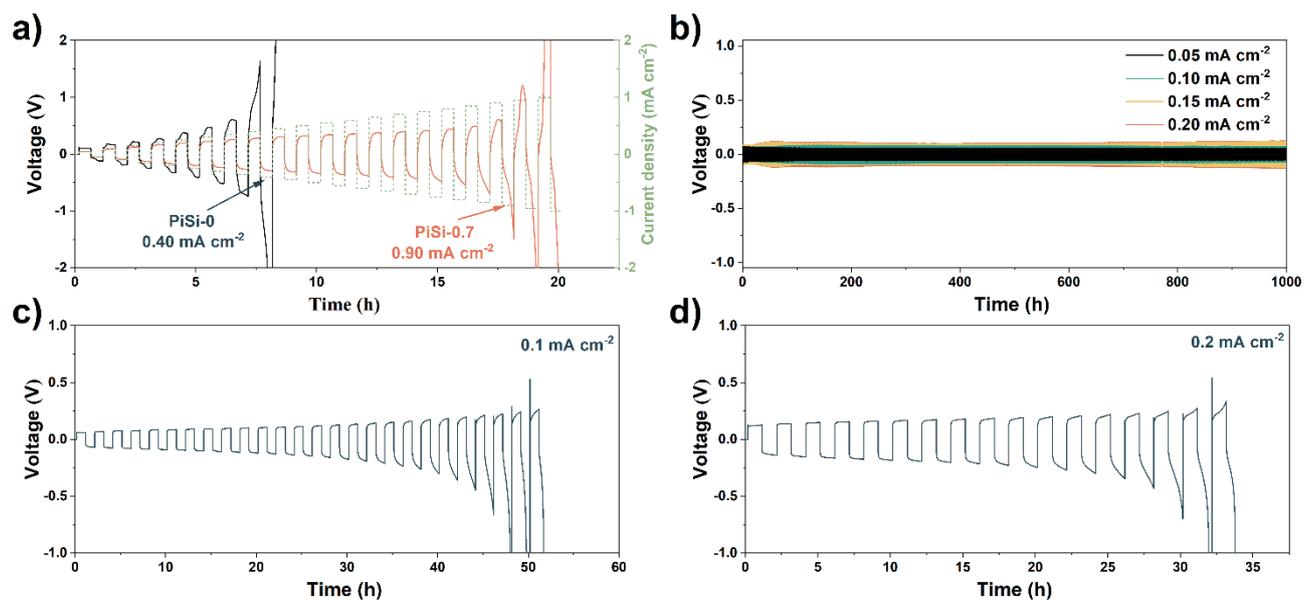

**Figure S13** (a) Radial distribution function (g(r)) of F and (b) Long-term cycling stability of Li||PiSi-0.7||Li symmetric cells at 25 °C. (c, d) Long-term cycling stability of Li||PiSi-0||Li symmetric cells at 25 °C.

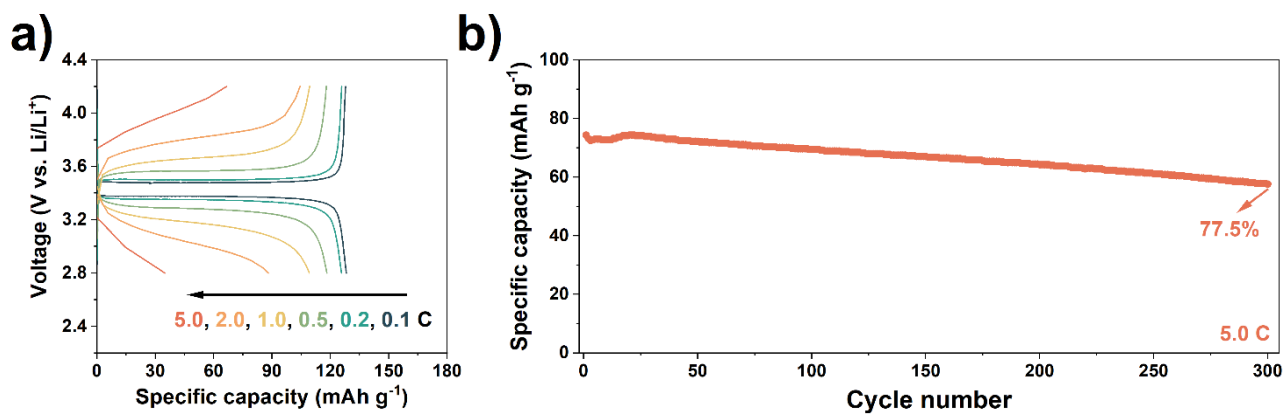

**Figure S14** Discharging-charging profiles of (a) LFP||PiSi-0||Li cell. (b) Long-term cycling performance of LFP||PiSi-0.7||Li at 5C at 25 °C.

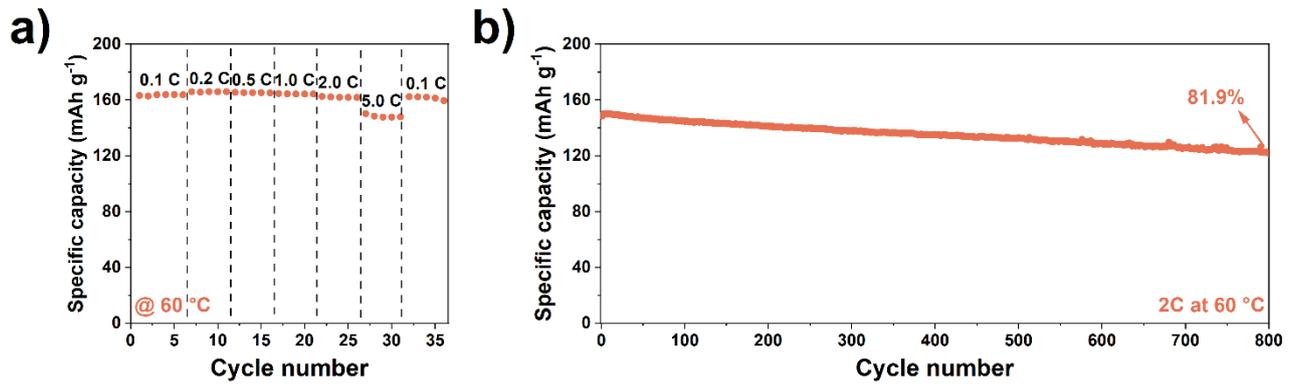

**Figure S15** (a) Rate capability of LFP||PiSi-0.7||Li cell at various current rates at 60 °C. (b) Long-term cycling performance of LFP||PiSi-0.7||Li at 2.0 C 60 °C.